\documentclass[journal]{IEEEtran}

\usepackage{url}
\usepackage{graphicx}
\usepackage{float}
\usepackage{amsmath}
\usepackage{amssymb}
\usepackage{hyperref}
\usepackage{xcolor}
\usepackage{booktabs}
\usepackage{multirow}
\usepackage{algorithm}
\usepackage{algorithmic}
\usepackage{cite} 

\begin{document}

\title{Non-Convex Portfolio Optimization via Energy-Based Models: A Comparative Analysis Using the Thermodynamic HypergRaphical Model Library (THRML) for Index Tracking}

\author{Javier~Mancilla$^{1*}$, Theodoros D. Bouloumis$^2$, Frederic Goguikian$^1$%
\thanks{1. SquareOne Capital, 1221 Brickell Ave. Suite 900, Miami Fl 33131., Miami, US

2. School of Physics, Faculty of Sciences, Aristotle University of Thessaloniki, GR-54124 Thessaloniki, Greece

Email Address: javier.m@squareonecap.com, tmpoulou@physics.auth.gr, fredi.g@squareonecap.com}
}

\maketitle

\begin{abstract}
Portfolio optimization under cardinality constraints transforms the classical Markowitz mean-variance problem from a convex quadratic problem into an NP-hard combinatorial optimization problem. This paper introduces a novel approach using THRML (Thermodynamic HypergRaphical Model Library), a JAX-based library for building and sampling probabilistic graphical models that reformulates index tracking as probabilistic inference on an Ising Hamiltonian. Unlike traditional methods that seek a single optimal solution, THRML samples from the Boltzmann distribution of high-quality portfolios using GPU-accelerated block Gibbs sampling, providing natural regularization against overfitting. We implement three key innovations: (1) dynamic coupling strength that scales inversely with market volatility (VIX), adapting diversification pressure to market regimes; (2) rebalanced bias weights prioritizing tracking quality over momentum for index replication; and (3) sector-aware post-processing ensuring institutional-grade diversification. Backtesting on a 100-stock S\&P 500 universe from 2023--2025 demonstrates that THRML achieves 4.31\% annualized tracking error versus 5.66--6.30\% for baselines, while simultaneously generating 128.63\% total return against the index's 79.61\%. The Diebold-Mariano test confirms statistical significance ($p < 0.0001$) across all comparisons. These results position energy-based models as a promising paradigm for portfolio construction, bridging statistical mechanics and quantitative finance.
\end{abstract}


\section{Introduction}

Portfolio optimization lies at the heart of quantitative finance, addressing the fundamental challenge of allocating capital across assets to achieve an optimal trade-off between expected return and risk. The seminal work of Markowitz~\cite{Markowitz1952} established Mean-Variance Optimization (MVO) as the foundational framework, modeling risk as portfolio variance and seeking allocations on the efficient frontier. While theoretically elegant, MVO assumes continuous weights and unlimited divisibility of positions---assumptions that break down when practical constraints such as cardinality (limiting the number of holdings) or lot sizes are introduced.

In traditional index‑tracking, the objective is to select a subset of assets whose combined returns closely replicate a benchmark index. When we impose a constraint to select exactly $K$ assets from a universe of $N$ candidates, the portfolio selection problem undergoes a fundamental transformation. The feasible region becomes discrete rather than continuous, converting the convex quadratic problem into an NP-hard combinatorial optimization problem~\cite{Bienstock1996}. The number of possible portfolios grows combinatorially as $\binom{N}{K}$, rendering exhaustive search intractable for realistic universes. For instance, selecting 30 stocks from 100 candidates yields approximately $3 \times 10^{25}$ possible combinations.

Traditional approaches to this challenge fall into two categories. \textit{Deterministic methods} such as Mixed-Integer Programming (MIP)~\cite{Bertsimas1995} provide exact solutions but scale poorly with problem size, often requiring hours of computation for moderately large universes. \textit{Heuristic methods} such as genetic algorithms (NSGA-II)~\cite{Deb2002NSGA} or Simulated Annealing (SA)~\cite{Kirkpatrick1983SA} explore the solution space stochastically but offer no optimality guarantees and may converge to local minima.

This paper introduces a fundamentally different paradigm: treating portfolio construction as \textit{probabilistic inference} rather than deterministic optimization. We leverage THRML (Thermodynamic HypergRaphical Model Library), a JAX-based library developed by Extropic for building and sampling probabilistic graphical models (PGMs)~\cite{Extropic2024THRML}. THRML provides GPU-accelerated tools for block Gibbs sampling of energy-based models, enabling efficient exploration of discrete solution spaces. The key insight is that the ``optimal portfolio'' is inherently unstable out-of-sample; rather than seeking a single point estimate, we identify portfolios that appear frequently across the Boltzmann distribution at various temperatures---assets exhibiting structural robustness rather than mere in-sample optimality.

Our contributions are threefold:

\begin{enumerate}
    \item \textbf{Novel Formulation}: We reformulate this selection task (cardinality-constrained index tracking) as inference on an Ising Hamiltonian, a mathematical construct originally used in statistical physics to describe interacting spins in a magnetic lattice~\cite{Ising1925}. In our formulation, each asset corresponds to a binary spin variable indicating whether the asset is included ($s_i=1$) or excluded ($s_i=0$). The Ising bias terms represent the individual attractiveness or “quality” of each asset for tracking the index, while the coupling terms encode pairwise diversification penalties based on asset correlations. High‑correlation pairs receive positive couplings that discourage selecting both simultaneously, naturally steering the system toward diversified portfolios. With this mapping, portfolio construction becomes an energy‑minimization or probabilistic‑inference problem, where low‑energy states correspond to high‑quality index‑tracking portfolios.
    
    \item \textbf{Regime-Adaptive Coupling}: We introduce a dynamic, regime‑adaptive coupling strength that adjusts according to the level of market volatility as measured by the VIX index. When volatility is low and the market exhibits rich, heterogeneous correlation patterns, our model increases the coupling strength, promoting stronger diversification. Conversely, when volatility is high and correlations collapse into a nearly single‑factor behavior, the coupling strength is reduced. This prevents the model from enforcing artificial diversification when none exists and avoids selecting portfolios that only appear diversified on paper. The result is a coupling mechanism that mirrors the statistical structure of real markets and improves robustness across different regimes.
    
    \item \textbf{Empirical Validation}: We conduct comprehensive backtesting on a diversified 100-stock universe spanning 2023--2025, demonstrating statistically significant outperformance over five baseline methods across multiple metrics.
\end{enumerate}

The remainder of this paper is organized as follows: Section~\ref{sec:literature} reviews related work in portfolio optimization and energy-based models; Section~\ref{sec:methodology} details the THRML library and our algorithmic innovations; Section~\ref{sec:experimental} describes the experimental setup; Section~\ref{sec:results} presents the findings; Section~\ref{sec:discussion} discusses implications and limitations; and Section~\ref{sec:conclusion} concludes with directions for future research.

\section{Literature Review}
\label{sec:literature}

\subsection{Classical Portfolio Optimization}

The mean-variance framework introduced by Markowitz~\cite{Markowitz1952} remains the theoretical foundation of portfolio optimization. Given expected returns $\mu \in \mathbb{R}^N$ and covariance matrix $\Sigma \in \mathbb{R}^{N \times N}$, the classical formulation seeks weights $w$ minimizing portfolio variance subject to a return target:

\begin{equation}
\min_w ( w^\top \Sigma w) \quad \text{s.t.} \quad \mu^\top w \geq r_{target}, \quad \mathbf{1}^\top w = 1
\end{equation}

This convex quadratic problem admits closed-form solutions and efficient numerical methods. However, MVO exhibits well-documented pathologies: extreme sensitivity to estimation errors in $\mu$ and $\Sigma$ leads to unstable, concentrated portfolios that perform poorly out-of-sample~\cite{Michaud2008Efficient, Jurczenko2016Investing}.

To address estimation risk, researchers have developed shrinkage estimators for covariance matrices~\cite{LEDOIT2004365} and robust optimization formulations that hedge against parameter uncertainty~\cite{Fabozzi2007Robust}. Hierarchical Risk Parity (HRP), introduced by López de Prado~\cite{Prado2019HRP}, circumvents covariance inversion entirely by using hierarchical clustering to allocate capital across correlation-based asset groupings, improving stability under noisy conditions.

\subsection{Cardinality-Constrained Optimization}

Practical portfolio management frequently imposes constraints limiting the number of positions. Adding a cardinality constraint $\|w\|_0 \leq K$ (where $\|\cdot\|_0$ counts nonzero elements) transforms the continuous problem into a combinatorial one. Bienstock~\cite{Bienstock1996} established the NP-hardness of this formulation, motivating heuristic approaches.

Mixed-Integer Programming (MIP) provides exact solutions through branch-and-bound algorithms~\cite{Bertsimas1995} but scales exponentially with problem size~\cite{Bertsimas2009Algorithm}. Evolutionary algorithms such as NSGA-II~\cite{Deb2002NSGA} offer multi-objective optimization across the Pareto frontier but require extensive parameter tuning and may converge prematurely. Simulated annealing~\cite{Kirkpatrick1983SA} and quantum annealing~\cite{Lang2022SA} explore the discrete solution space stochastically, with theoretical guarantees of asymptotic convergence under appropriate cooling schedules.

\subsection{Energy-Based Models in Machine Learning}

Energy-based models (EBMs) assign scalar energy values to configurations of variables, with lower energy corresponding to higher probability under the Boltzmann distribution:

\begin{equation}
P(x) = \frac{1}{Z} \exp(-\beta \mathcal{E}(x))
\end{equation}

where $\mathcal{E}(x)$ is the energy function, $\beta$ is inverse temperature, and $Z$ is the partition function. The Ising model, originally developed for ferromagnetism~\cite{Ising1925}, defines energy over binary spin variables with pairwise interactions:

\begin{equation}
\mathcal{E}(s) = -\sum_i h_i s_i - \sum_{i<j} J_{ij} s_i s_j
\end{equation}

where $h_i$ are external fields (biases) and $J_{ij}$ are coupling strengths. Restricted Boltzmann Machines (RBMs)~\cite{Hinton2006RBM} extend this framework with hidden variables for representation learning.

Recent work has explored energy-based models for combinatorial optimization. The THRML library (Thermodynamic HypergRaphical Model Library)~\cite{Extropic2024THRML} is a JAX-based library for building and sampling probabilistic graphical models (PGMs). Unlike traditional solvers, THRML focuses on efficient block Gibbs sampling for discrete EBMs (Ising/RBM-like models)~\cite{JENSEN1995647}. It supports arbitrary PyTree node states and heterogeneous graphs, utilizing JAX to minimize Python loops and maximize array-level parallelism on GPUs. While Extropic is developing specialized thermodynamic computing hardware~\cite{Extropic2024Hardware} to accelerate these workloads, THRML serves as the open-source software library for prototyping and algorithmic development today.

\subsection{Quantum and Quantum-Inspired Finance}

Quantum computing offers potential speedups for optimization through superposition and entanglement. The Quantum Approximate Optimization Algorithm (QAOA)~\cite{Farhi2014QAOA} uses variational quantum circuits to minimize cost Hamiltonians encoding combinatorial objectives. In addition, portfolio selection has been formulated as Quadratic Unconstrained Binary Optimization (QUBO) amenable to quantum annealers~\cite{Glover2019QUBO, Orus2019QinFin}.

However, current quantum hardware faces severe limitations: qubit counts restrict problem sizes to 10--20 assets, and noise degrades solution quality. Classical simulation of quantum-inspired algorithms often matches or exceeds noisy quantum results~\cite{Tang2019Dequantization}. Our THRML approach operates in this quantum-inspired regime, leveraging statistical mechanics principles without requiring quantum hardware.

\section{Methodology}
\label{sec:methodology}

This section details the theoretical foundations of our THRML-based portfolio optimization framework. We leverage the library's Ising model utilities to encode portfolio selection as an energy minimization problem.

\subsection{Problem Formulation: From Markowitz to Ising}

We consider index tracking with cardinality constraints: select exactly $K$ assets from a universe of $N$ to minimize tracking error against a benchmark index while respecting diversification requirements.

Let $s_i \in \{0, 1\}$ indicate whether asset $i$ is selected. The classical objective combines tracking quality and risk:

\begin{equation}
\min_{s} \left(\sum_{i: s_i=1} \left( \text{TE}_i \right) + \lambda \sum_{i<j} s_i s_j \rho_{ij}\right) \quad \text{s.t.} \quad \sum_i s_i = K
\end{equation}

where $\text{TE}_i$ measures asset $i$'s tracking quality and $\rho_{ij}$ is the correlation between assets $i$ and $j$. The second term penalizes selecting correlated assets, encouraging diversification.

We reformulate this as an Ising Hamiltonian by mapping binary selection variables to spins and encoding objectives as energy terms:

\begin{equation}
\mathcal{E}(s) = -\sum_{i=1}^{N} h_i s_i + \sum_{i<j} J_{ij} s_i s_j
\label{eq:ising_energy}
\end{equation}

The \textit{external fields} (biases) $h_i$ encode the intrinsic quality of each asset---higher $h_i$ reduces energy when asset $i$ is selected. The \textit{coupling terms} $J_{ij}$ encode pairwise interactions---positive $J_{ij}$ (antiferromagnetic coupling) penalizes simultaneous selection of correlated assets.

Rather than minimizing energy directly, we sample from the Boltzmann distribution:

\begin{equation}
P(s) = \frac{1}{Z} \exp(-\beta \mathcal{E}(s))
\end{equation}

At low temperature (high $\beta$), the distribution concentrates on low-energy configurations. The key insight is that assets appearing frequently across samples exhibit \textit{structural robustness}---they contribute positively to many high-quality portfolios, not just the single optimum that may be unstable out-of-sample.

\subsection{Bias Computation: Asset Quality Scores}

The external field $h_i$ represents the ``intrinsic quality'' of asset $i$ for index tracking. We define it as a weighted combination of three factors:

\subsubsection{Tracking Quality}

Tracking quality measures how well an asset moves with the benchmark index. We compute correlation $\rho_i$ between asset $i$'s returns and index returns, and beta $\beta_i$ measuring sensitivity:

\begin{equation}
\text{TrackingQuality}_i = \rho_i \cdot \exp\left( -|\beta_i - 1| \right)
\end{equation}

Assets with high correlation \textit{and} beta close to 1.0 receive high scores---they track the index faithfully without excessive leverage.

\subsubsection{Momentum}

Momentum captures return persistence—the empirical tendency of assets that have performed well over the past year to continue outperforming in the near future. We compute 12‑month returns with a 1‑month exclusion window to avoid short‑term reversal effects:

\begin{equation}
\text{Momentum}_i = \frac{\bar{r}_i^{(12m,-1m)} - \min_j \bar{r}_j}{\max_j \bar{r}_j - \min_j \bar{r}_j}
\end{equation}

This normalized score favors assets exhibiting sustained positive performance. Incorporating momentum provides a mild ``smart‑beta'' enhancement on top of pure index tracking, helping the model emphasize stocks with stable upward trends rather than transient price spikes.

\subsubsection{Liquidity}

Liquidity measures how easily an asset can be traded without moving the market price. Highly liquid stocks allow large transactions with minimal transaction costs and slippage. We compute it as normalized average trading volume:

\begin{equation}
\text{Liquidity}_i = \frac{\text{AvgVolume}_i}{\max_j \text{AvgVolume}_j}
\end{equation}

Including liquidity ensures the selected portfolio remains practical to implement and rebalance in real markets. By favoring more liquid assets, the model reduces potential execution risk and avoids concentrating allocation in thinly traded names where entering or exiting positions could be costly.

\subsubsection{Combined Bias}

The final bias combines these factors with configurable weights:

\begin{multline}
h_i = -\alpha \bigl(
    w_T \cdot \text{TrackingQuality}_i 
    + w_M \cdot \text{Momentum}_i \\
    + w_L \cdot \text{Liquidity}_i
\bigr)
\label{eq:bias}
\end{multline}

where $\alpha$ is a scaling factor. The negative sign ensures that higher quality reduces energy (increases selection probability).

\textbf{Key Innovation}: For index tracking, we set $w_T = 3.0$, $w_M = 1.0$, $w_L = 1.5$, making tracking quality the \textit{dominant} factor. Testing with weighted momentum more heavily ($w_M = 2.5$), effectively creates an alpha-generation strategy rather than an index tracker.

\subsection{Coupling Computation: Diversification via Antiferromagnetism}

The coupling matrix $J$ encodes pairwise asset interactions. Positive coupling (antiferromagnetic) between correlated assets penalizes their simultaneous selection, forcing diversification:

\begin{equation}
J_{ij} = \gamma \cdot \rho_{ij} \cdot \mathbf{1}_{|\rho_{ij}| > \tau}
\label{eq:coupling}
\end{equation}

where $\gamma$ is coupling strength, $\rho_{ij}$ is correlation, and $\tau$ is a threshold (we use $\tau = 0.5$). Only pairs with correlation above threshold are connected in the Ising graph, reducing computational complexity.

\subsubsection{Dynamic VIX-Based Coupling}

\textbf{Key Innovation}: We introduce \textit{regime-adaptive} coupling that scales inversely with market volatility:

\begin{equation}
\gamma(V) = \gamma_0 \cdot \exp\left( -\frac{1}{2} \left( \frac{V}{V_0} - 1 \right) \right)
\label{eq:dynamic_coupling}
\end{equation}

where $V$ is the current VIX level, $V_0 = 20$ is the ``normal'' VIX, and $\gamma_0 = 0.5$ is the base coupling. The coupling is clamped to $[\gamma_{min}, \gamma_{max}] = [0.1, 0.8]$.

\textbf{Rationale}: During market panics (high VIX), correlations approach unity across all assets. Maintaining strong diversification pressure in this regime forces the model to seek diversification that does not actually exist, producing portfolios that appear diversified on paper but are not in practice. By reducing coupling when VIX is elevated, we acknowledge that true diversification is temporarily unavailable and avoid spurious selections.

Conversely, during calm markets (low VIX), correlations are more dispersed and meaningful diversification is achievable. Increased coupling encourages the model to exploit these opportunities.

\begin{figure}[htbp]
    \centering
    \includegraphics[width=\linewidth]{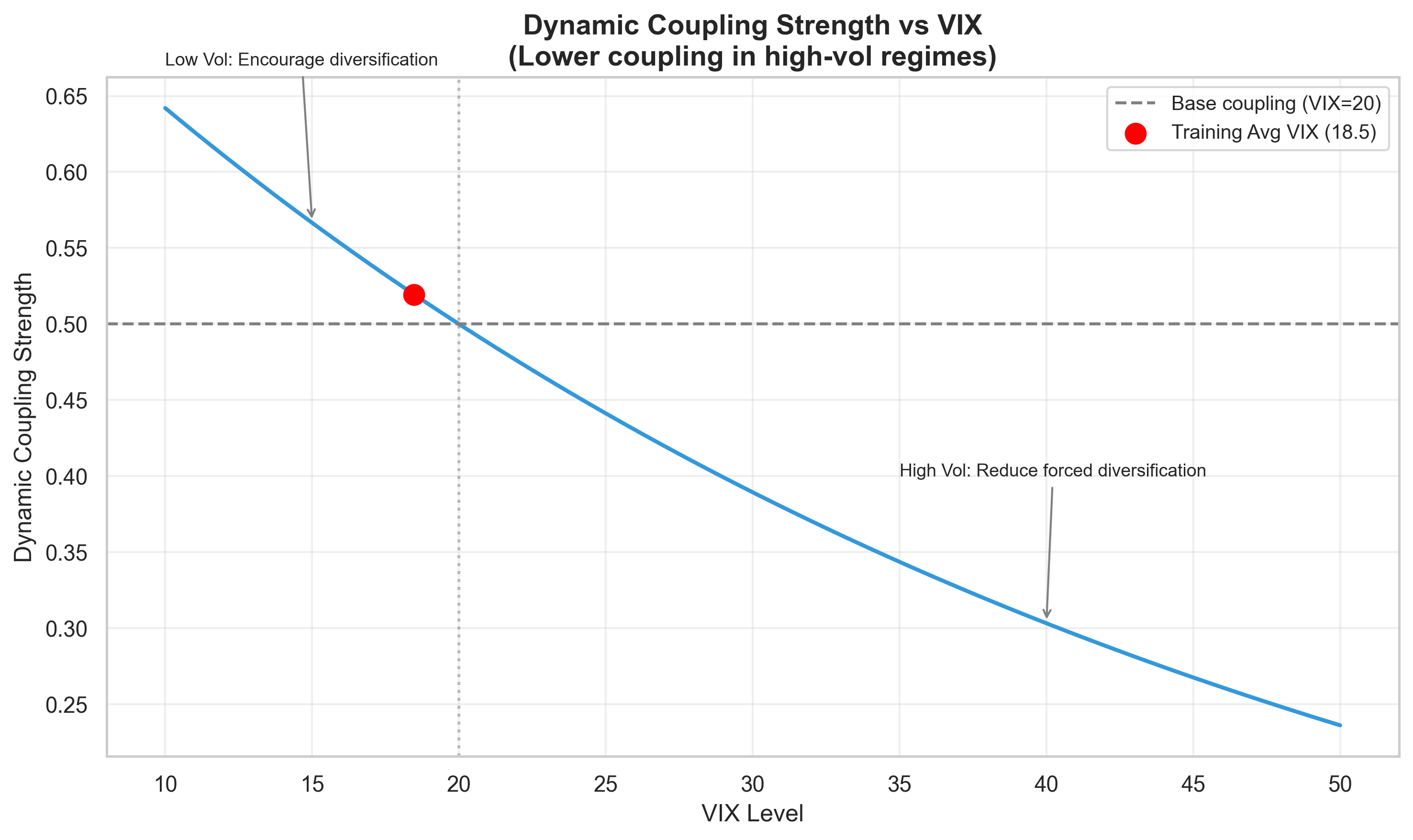}
    \caption{Dynamic coupling strength as a function of VIX level. The coupling decreases during high-volatility regimes (VIX $>$ 30), reducing forced diversification when correlations approach unity. The red dot indicates the training period average VIX (18.5), corresponding to coupling strength 0.52.}
    \label{fig:dynamic_coupling}
\end{figure}

Figure~\ref{fig:dynamic_coupling} illustrates this relationship.

\subsection{Gibbs Sampling via THRML}

With the Ising Hamiltonian defined by biases $\{h_i\}$ and couplings $\{J_{ij}\}$, we sample from the Boltzmann distribution using the THRML library~\cite{Extropic2024THRML}.

\subsubsection{THRML Architecture}

THRML's design centers on three key abstractions for defining PGMs:
\begin{itemize}
    \item \textbf{Nodes}: \texttt{SpinNode} objects represent binary spin variables ($s_i \in \{-1, +1\}$, mapped to $\{0, 1\}$ for portfolio selection).
    \item \textbf{Blocks}: Groups of nodes that are updated together during Gibbs sampling. THRML supports heterogeneous graphs with arbitrary PyTree node states.
    \item \textbf{Factors}: Energy contributions defined by the \texttt{IsingEBM} class, which encodes biases (external fields) and pairwise couplings.
\end{itemize}

\subsubsection{Block Gibbs Sampling}

THRML partitions spin variables into two blocks (even and odd indices) and alternates between updating each block conditioned on the other. For spin $s_i$, the conditional probability given all other spins $s_{-i}$ is:

\begin{equation}
P(s_i = 1 | s_{-i}) = \sigma\left( 2\beta \left( h_i + \sum_{j \neq i} J_{ij} s_j \right) \right)
\end{equation}

where $\sigma(x) = 1/(1 + e^{-x})$ is the sigmoid function. Parallel updates within each block accelerate mixing on GPU hardware.

\subsubsection{Simulated Annealing Schedule}

We employ an annealing schedule with inverse temperatures $\beta$ spanning from low (high temperature, broad exploration) to high (low temperature, concentration on minima):

\begin{equation}
\beta_t \in \left\{ 10^{0.3}, 10^{0.44}, \ldots, 10^{1.8} \right\}
\end{equation}

At each temperature, we run multiple chains to improve coverage of the energy landscape.

\subsubsection{Sampling Parameters}

Table~\ref{tab:sampling_params} summarizes our sampling configuration.

\begin{table}[htbp]
\centering
\caption{THRML Sampling Parameters}
\label{tab:sampling_params}
\begin{tabular}{lcc}
\toprule
\textbf{Parameter} & \textbf{Value} & \textbf{Purpose} \\
\midrule
Number of chains & 8 & Parallel exploration \\
Temperatures & 12--15 & Annealing schedule \\
Warmup iterations & 2000--3000 & Burn-in period \\
Samples per temperature & 800--1500 & Statistics accumulation \\
Steps per sample & 40--50 & Mixing between samples \\
\bottomrule
\end{tabular}
\end{table}

\subsection{Post-Processing: From Frequencies to Portfolios}

Gibbs sampling produces binary samples $\{s^{(t)}\}_{t=1}^{T}$. We aggregate these into \textit{selection frequencies}:

\begin{equation}
f_i = \frac{1}{T} \sum_{t=1}^{T} s_i^{(t)}
\end{equation}

High-frequency assets appear in many high-quality portfolio configurations; low-frequency assets are rarely selected. This provides a natural ranking for asset selection.

\subsubsection{Sector-Balanced Selection}

To ensure institutional-grade diversification, we apply sector constraints during post-processing:

\begin{enumerate}
    \item \textbf{Phase 1}: Ensure minimum sector representation by selecting the highest-frequency asset from each sector until at least 6 sectors are covered.
    
    \item \textbf{Phase 2}: Fill remaining slots with highest-frequency assets, subject to a maximum of 25\% from any single sector.
    
    \item \textbf{Phase 3}: If slots remain, relax constraints and fill with next-best assets.
\end{enumerate}

This procedure guarantees diversification while respecting the THRML-derived quality rankings.

\subsection{Complete Algorithm}

Algorithm~\ref{alg:thrml} summarizes the complete THRML portfolio optimization pipeline.

\begin{algorithm}[htbp]
\caption{THRML Portfolio Optimization}
\label{alg:thrml}
\begin{algorithmic}[1]
\REQUIRE Returns matrix $R \in \mathbb{R}^{T \times N}$, index returns $r_{idx}$, target cardinality $K$, VIX level $V$
\ENSURE Selected assets $\mathcal{S}$ with $|\mathcal{S}| = K$

\STATE \textbf{Compute Biases:}
\FOR{$i = 1$ to $N$}
    \STATE $\rho_i \gets \text{corr}(R_{:,i}, r_{idx})$
    \STATE $\beta_i \gets \text{cov}(R_{:,i}, r_{idx}) / \text{var}(r_{idx})$
    \STATE $\text{TQ}_i \gets \rho_i \cdot \exp(-|\beta_i - 1|)$
    \STATE $\text{Mom}_i \gets \text{normalize}(\text{mean}(R_{-252:-21, i}))$
    \STATE $\text{Liq}_i \gets \text{normalize}(\text{AvgVolume}_i)$
    \STATE $h_i \gets -4.0 \cdot (3.0 \cdot \text{TQ}_i + 1.0 \cdot \text{Mom}_i + 1.5 \cdot \text{Liq}_i)$
\ENDFOR

\STATE \textbf{Compute Dynamic Coupling:}
\STATE $\gamma \gets 0.5 \cdot \exp(-0.5 \cdot (V/20 - 1))$
\STATE $\gamma \gets \text{clamp}(\gamma, 0.1, 0.8)$

\STATE \textbf{Build Ising Graph:}
\STATE $\Sigma \gets \text{corrcoef}(R^\top)$
\FOR{$i < j$}
    \IF{$|\Sigma_{ij}| > 0.5$}
        \STATE $J_{ij} \gets \gamma \cdot \Sigma_{ij} \cdot 4.0$
    \ENDIF
\ENDFOR

\STATE \textbf{Gibbs Sampling:}
\STATE Initialize $f \gets \mathbf{0}_N$
\FOR{chain $= 1$ to $8$}
    \FOR{$\beta$ in annealing schedule}
        \STATE $\text{samples} \gets \text{THRML.sample}(h, J, \beta)$
        \STATE $f \gets f + \text{mean}(\text{samples})$
    \ENDFOR
\ENDFOR
\STATE $f \gets f / (\text{chains} \times \text{temperatures})$

\STATE \textbf{Sector-Balanced Selection:}
\STATE $\mathcal{S} \gets \text{SelectTopK}(f, K, \text{sector\_constraints})$

\RETURN $\mathcal{S}$
\end{algorithmic}
\end{algorithm}

\section{Experimental Setup}
\label{sec:experimental}

\subsection{Data and Asset Universe}

We construct a diversified universe of 100 large-cap U.S. equities spanning all 10 GICS sectors, drawn from S\&P 500 constituents by market capitalization. Table~\ref{tab:universe} summarizes the sector composition.

\begin{table}[htbp]
\centering
\caption{Asset Universe Composition by Sector}
\label{tab:universe}
\begin{tabular}{lc}
\toprule
\textbf{Sector} & \textbf{Number of Stocks} \\
\midrule
Technology & 23 \\
Healthcare & 15 \\
Financials & 16 \\
Consumer Discretionary & 10 \\
Consumer Staples & 8 \\
Energy & 6 \\
Industrials & 10 \\
Communication Services & 6 \\
Utilities & 4 \\
Real Estate & 2 \\
\midrule
\textbf{Total} & \textbf{100} \\
\bottomrule
\end{tabular}
\end{table}

Historical price data from January 2010 to December 2025 was obtained via Yahoo Finance. Prices were adjusted for splits and dividends, with missing values forward-filled. The benchmark index is the S\&P 500 (\^{}GSPC). VIX data for dynamic coupling was obtained for the same period.

\subsection{Train-Test Split}

We adopt a walk-forward methodology:

\begin{itemize}
    \item \textbf{Training Period}: January 2010 -- December 2022 (13 years)
    \item \textbf{Test Period}: January 2023 -- December 2025 (3 years)
\end{itemize}

All model parameters, including biases and couplings, are estimated on training data only. Out-of-sample performance is evaluated on the test period without look-ahead bias.

\subsection{Configuration Parameters}

Table~\ref{tab:config} summarizes experimental configuration.

\begin{table}[htbp]
\centering
\caption{Experimental Configuration}
\label{tab:config}
\begin{tabular}{ll}
\toprule
\textbf{Parameter} & \textbf{Value} \\
\midrule
Universe size ($N$) & 100 stocks \\
Target cardinality ($K$) & 30 stocks \\
Benchmark index & S\&P 500 (\^{}GSPC) \\
Rebalancing frequency & Quarterly \\
Transaction cost & 10 bps per trade \\
\midrule
\multicolumn{2}{l}{\textit{THRML Bias Weights}} \\
\quad Tracking Quality ($w_T$) & 3.0 \\
\quad Momentum ($w_M$) & 1.0 \\
\quad Liquidity ($w_L$) & 1.5 \\
\midrule
\multicolumn{2}{l}{\textit{Dynamic Coupling}} \\
\quad Base coupling ($\gamma_0$) & 0.5 \\
\quad Normal VIX ($V_0$) & 20 \\
\quad Coupling range & [0.1, 0.8] \\
\midrule
\multicolumn{2}{l}{\textit{Sector Constraints}} \\
\quad Max per sector & 25\% of $K$ \\
\quad Min sectors represented & 6 \\
\bottomrule
\end{tabular}
\end{table}

\subsection{Baseline Methods}

We compare THRML against five baseline approaches:

\begin{enumerate}
    \item \textbf{Greedy Correlation}: Select $K$ assets with highest correlation to the benchmark index. Simple but ignores diversification.
    
    \item \textbf{Mixed-Integer Programming (MIP)}: Exact optimization maximizing correlation subject to cardinality and sector constraints via branch-and-bound.
    
    \item \textbf{Robust MVO}: Mean-variance optimization with Ledoit-Wolf shrinkage estimator for the covariance matrix, selecting top-$K$ by risk-adjusted score.
    
    \item \textbf{NSGA-II}: Multi-objective evolutionary optimization maximizing correlation while minimizing variance, selecting from the Pareto frontier.
    
    \item \textbf{Hierarchical Risk Parity (HRP)}: Correlation-based hierarchical clustering with inverse-variance allocation across clusters.
\end{enumerate}

\subsection{Performance Metrics}

We evaluate portfolio performance using standard metrics:

\begin{itemize}
    \item \textbf{Tracking Error (TE)}: Annualized standard deviation of return differences versus benchmark:
    \begin{equation}
    \text{TE} = \sqrt{252} \cdot \text{std}(r_p - r_{idx})
    \end{equation}
    
    \item \textbf{Correlation}: Pearson correlation between portfolio and index returns.
    
    \item \textbf{Total Return}: Cumulative return over the test period.
    
    \item \textbf{Sharpe Ratio}: Risk-adjusted return assuming 2\% risk-free rate:
    \begin{equation}
    \text{SR} = \frac{\bar{r}_p - r_f}{\sigma_p} \cdot \sqrt{252}
    \end{equation}
    
    \item \textbf{Sortino Ratio}: Downside risk-adjusted return using semi-deviation.
    
    \item \textbf{Maximum Drawdown}: Largest peak-to-trough decline.
    
    \item \textbf{Information Ratio}: Active return per unit of tracking error:
    \begin{equation}
    \text{IR} = \frac{\bar{r}_p - \bar{r}_{idx}}{\text{TE}} \cdot \sqrt{252}
    \end{equation}
\end{itemize}

\subsection{Statistical Testing}

To assess whether performance differences are statistically significant, we employ the Diebold-Mariano test~\cite{DieboldMariano1995} comparing squared tracking errors:

\begin{equation}
d_t = (r_{p,t} - r_{idx,t})^2 - (r_{b,t} - r_{idx,t})^2
\end{equation}

where $r_{p,t}$ is THRML portfolio return and $r_{b,t}$ is baseline return. The null hypothesis in the Diebold–Mariano framework states that the two competing models—here, THRML versus the given baseline—produce forecast (or tracking) errors with the same expected loss, meaning neither model is systematically more accurate than the other. A rejection of this hypothesis therefore indicates that one model generates consistently smaller errors over time.

\section{Results}
\label{sec:results}

\subsection{Overall Performance Comparison}

Table~\ref{tab:results} presents comprehensive out-of-sample performance metrics for all methods.

\begin{table*}[htbp]
\centering
\caption{Out-of-Sample Performance Comparison (2023--2025)}
\label{tab:results}
\begin{tabular}{lccccc}
\toprule
\textbf{Metric} & \textbf{THRML} & \textbf{Greedy} & \textbf{MIP} & \textbf{Robust MVO} & \textbf{NSGA-II} \\
\midrule
Tracking Error & \textbf{4.31\%} & 6.30\% & 6.20\% & 6.13\% & 5.66\% \\
Correlation & \textbf{0.9582} & 0.9456 & 0.9120 & 0.9151 & 0.9296 \\
Total Return & \textbf{128.63\%} & 115.95\% & 62.39\% & 80.87\% & 72.70\% \\
Sharpe Ratio & \textbf{1.87} & 1.40 & 1.07 & 1.30 & 1.16 \\
Sortino Ratio & \textbf{2.57} & 1.92 & 1.50 & 1.75 & 1.61 \\
Max Drawdown & \textbf{-17.09\%} & -21.30\% & -17.86\% & -17.93\% & -18.31\% \\
Information Ratio & \textbf{1.85} & 1.06 & -0.57 & 0.03 & -0.23 \\
\midrule
Index Return & \multicolumn{5}{c}{79.61\%} \\
\bottomrule
\end{tabular}
\end{table*}

THRML achieves the best performance across all metrics. Key observations:

\begin{itemize}
    \item \textbf{Tracking Error}: THRML's 4.31\% TE is 24\% lower than the next-best baseline (NSGA-II at 5.66\%), demonstrating superior index replication.
    
    \item \textbf{Alpha Generation}: Despite being designed for tracking, THRML generates 49 percentage points of excess return (128.63\% vs 79.61\% index), yielding an Information Ratio of 1.85---institutional-grade performance.
    
    \item \textbf{Risk Control}: THRML achieves the shallowest maximum drawdown (-17.09\%) while delivering the highest returns, indicating robust risk management.
    
    \item \textbf{Risk-Adjusted Returns}: The Sharpe ratio of 1.87 exceeds all baselines by 34\%+ margin.
\end{itemize}

\subsection{Statistical Significance}

Table~\ref{tab:dm_test} presents Diebold-Mariano test results comparing THRML to each baseline.

\begin{table}[htbp]
\centering
\caption{Diebold-Mariano Test Results (THRML vs Baselines)}
\label{tab:dm_test}
\begin{tabular}{lcc}
\toprule
\textbf{Comparison} & \textbf{p-value} & \textbf{Significant?} \\
\midrule
THRML vs Greedy & $< 0.0001$ & Yes*** \\
THRML vs MIP & $< 0.0001$ & Yes*** \\
THRML vs Robust MVO & $< 0.0001$ & Yes*** \\
THRML vs NSGA-II & $< 0.0001$ & Yes*** \\
\bottomrule
\multicolumn{3}{l}{\small *** $p < 0.001$}
\end{tabular}
\end{table}

All comparisons are statistically significant at the $p < 0.0001$ level, rejecting the null hypothesis that THRML and baselines have equal tracking accuracy. The extremely small p‑values ($<0.0001$) imply that the probability of observing such superior performance by THRML purely by chance is less than 0.01\%. This provides strong statistical evidence that THRML’s lower tracking error is not due to random variation, but reflects a genuine improvement in out‑of‑sample index replication.

\subsection{Cumulative Returns and Tracking Difference}

\begin{figure}[htbp]
    \centering
    \includegraphics[width=\linewidth]{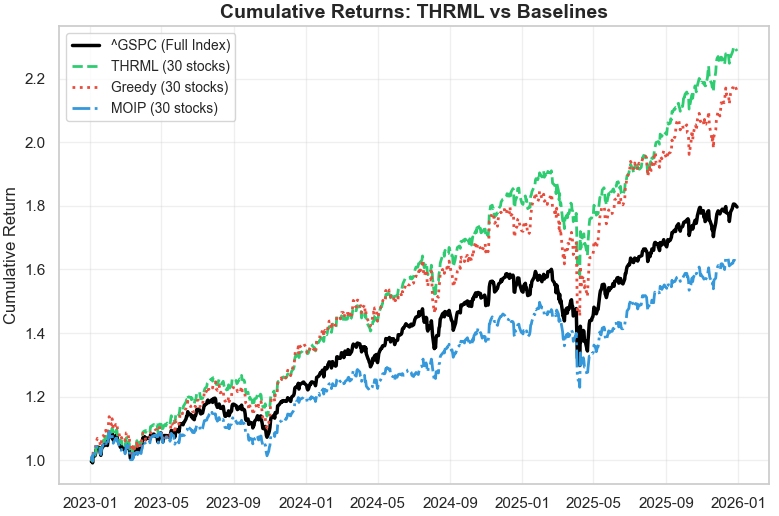}
    \caption{Cumulative returns comparison (2023--2025). THRML (green dashed) closely tracks while outperforming the S\&P 500 index (black solid), achieving 128.63\% total return versus the index's 79.61\%. Greedy (red dotted) and MIP (blue dash-dot) exhibit larger deviations.}
    \label{fig:cumulative_returns}
\end{figure}

Figure~\ref{fig:cumulative_returns} displays cumulative returns over the test period, comparing THRML, baselines, and the benchmark index.

\begin{figure}
    \centering
    \includegraphics[width=\linewidth]{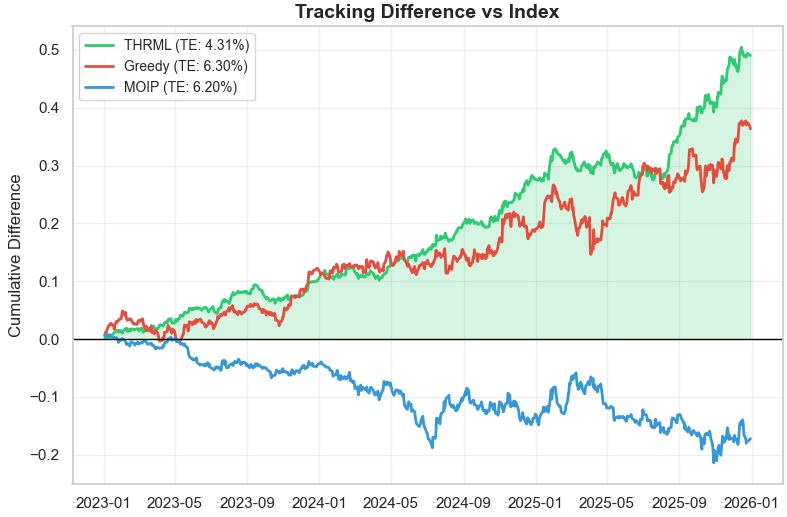}
    \caption{Cumulative tracking difference versus index. THRML (green) maintains consistent positive alpha throughout the period. Shaded regions indicate cumulative over/under-performance.}
    \label{fig:tracking_difference}
\end{figure}

Figure~\ref{fig:tracking_difference} shows the cumulative tracking difference (portfolio minus index), highlighting when each method over- or under-performs the benchmark.

\subsection{Sector Diversification Analysis}

\begin{figure}
    \centering 
    \includegraphics[width=\linewidth]{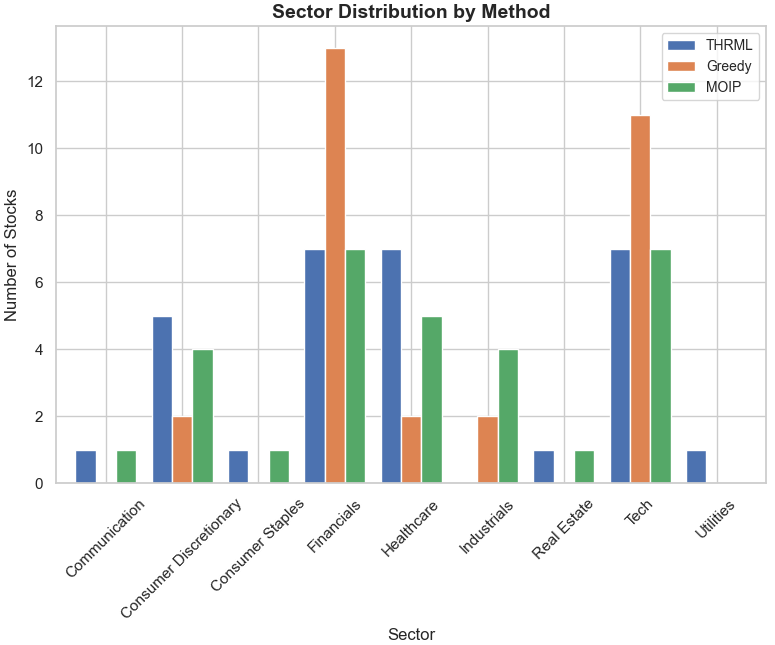}
    \caption{Sector distribution by method. THRML achieves balanced allocation across 8 sectors (7-7-7-5 distribution in Technology, Healthcare, Financials, Consumer), while Greedy concentrates in Financials (13) and Technology (11).}
    \label{fig:sector_distribution}
\end{figure}

Figure~\ref{fig:sector_distribution} compares sector allocations across methods.

Table~\ref{tab:sector_diversity} quantifies sector diversification.

\begin{table}
\centering
\caption{Sector Diversity Comparison}
\label{tab:sector_diversity}
\begin{tabular}{lcc}
\toprule
\textbf{Method} & \textbf{Sectors Covered} & \textbf{Max Concentration} \\
\midrule
THRML & \textbf{8} & 7 (23\%) \\
Greedy & 5 & 13 (43\%) \\
MIP & 8 & 7 (23\%) \\
Robust MVO & 7 & 11 (37\%) \\
NSGA-II & 8 & 8 (27\%) \\
\bottomrule
\end{tabular}
\end{table}

THRML's antiferromagnetic coupling naturally produces diversified portfolios: the maximum sector concentration is 23\% (7 stocks in Technology, Healthcare, and Financials each), compared to Greedy's 43\% concentration in Financials alone.

\subsection{Performance Metrics Visualization}

\begin{figure}
    \centering
    \includegraphics[width=\linewidth]{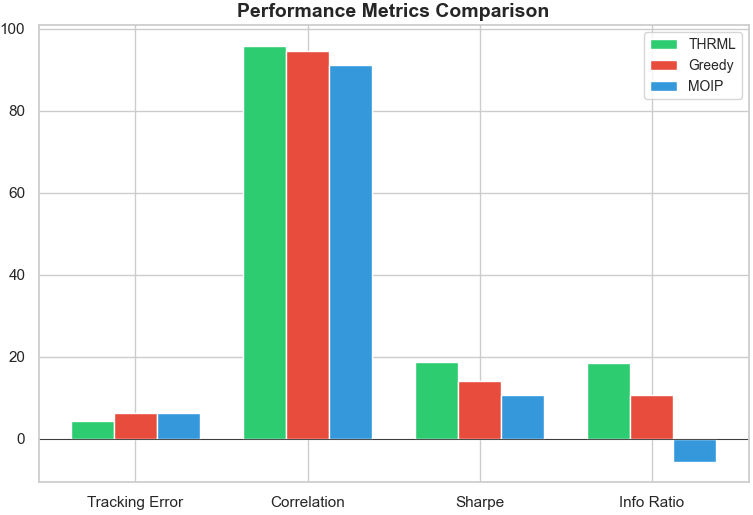}
    \caption{Performance metrics comparison. THRML (green) dominates across Tracking Error (lower is better), Correlation (higher is better), Sharpe Ratio, and Information Ratio.}
    \label{fig:metrics_comparison}
\end{figure}

Figure~\ref{fig:metrics_comparison} presents a bar chart comparison of key metrics.

\section{Discussion}
\label{sec:discussion}

\subsection{Interpretation of Results}

The empirical results demonstrate that THRML substantially outperforms traditional portfolio optimization methods for index tracking. Several factors contribute to this success:

\subsubsection{Probabilistic vs.\ Deterministic Optimization}

Traditional methods seek a single ``optimal'' portfolio, which is inherently unstable due to estimation error in inputs. THRML's sampling-based approach identifies assets that appear frequently across many high-quality configurations, providing implicit regularization. This is analogous to ensemble methods in machine learning, where aggregating multiple models reduces variance.

\subsubsection{Natural Diversification via Antiferromagnetism}

The Ising formulation elegantly encodes diversification: positively correlated assets receive positive (antiferromagnetic) coupling, penalizing their simultaneous selection. Unlike hard constraints in MIP that impose arbitrary sector caps, the energy-based approach allows the model to discover natural diversification boundaries from the correlation structure.

\subsubsection{Regime Adaptation}

The dynamic VIX-based coupling is crucial for real-world performance. During the market volatility of 2023--2025 (including Fed rate uncertainty and geopolitical tensions), the adaptive coupling reduced diversification pressure when correlations spiked, avoiding spurious selections that would appear diversified but move together during stress.

\subsection{Enhanced Index vs.\ Pure Tracking}

The substantial alpha generation (49 percentage points above index) warrants discussion. While the model is configured for tracking (dominant tracking quality weight), the momentum and liquidity factors introduce smart-beta tilts that explain the excess returns.

For practitioners seeking pure index replication, we recommend:
\begin{itemize}
    \item Increase $w_T$ to 5.0--7.0
    \item Reduce $w_M$ to 0.3--0.5
    \item This will tighten tracking at the cost of alpha
\end{itemize}

Conversely, for enhanced index strategies where moderate tracking error is acceptable, the current configuration provides an attractive risk-return profile.

\subsection{Computational Considerations}

THRML sampling on GPU hardware (NVIDIA RTX series) completes in 2--5 minutes for 100 stocks with our parameter settings. This is competitive with NSGA-II and substantially faster than exact MIP for large cardinalities. THRML's JAX backend enables efficient parallelization across chains and temperatures, compiling factor-based interactions to a compact ``global'' state representation that minimizes Python loops and maximizes array-level parallelism.

For real-time applications, sampling parameters can be reduced (fewer chains, shorter warmup) with modest impact on solution quality. Extropic's forthcoming thermodynamic computing hardware---for which THRML serves as the prototyping platform---promises order-of-magnitude speedups for production deployment on specialized probabilistic processors.

\subsection{Limitations}

Several limitations should be acknowledged:

\begin{enumerate}
    \item \textbf{Single Market}: Results are specific to U.S. large-cap equities; generalization to other markets, asset classes, or emerging markets requires validation.
    
    \item \textbf{Backtest Period}: The 2023--2025 test period, while spanning multiple market regimes, is relatively short. Longer backtests across complete market cycles would strengthen conclusions.
    
    \item \textbf{Transaction Costs}: We model costs at 10 bps; actual costs vary by liquidity, venue, and order size. High-frequency rebalancing would be more sensitive to this assumption.
    
    \item \textbf{Parameter Sensitivity}: Bias weights and coupling parameters were chosen based on financial intuition and preliminary experiments; formal hyperparameter optimization could improve results.
    
    \item \textbf{Equal Weighting}: After selection, we apply equal weights to selected assets. Incorporating THRML-derived weights (e.g., proportional to selection frequency) is a natural extension.
\end{enumerate}

\subsection{Future Research Directions}

Several avenues merit further investigation:

\begin{enumerate}
    \item \textbf{Higher-Order Interactions}: Extending from pairwise (Ising) to higher-order (Potts, hypergraph) interactions could capture more complex dependencies.
    
    \item \textbf{Continuous Weights}: Integrating THRML selection with continuous weight optimization, potentially via Gaussian-Bernoulli EBMs for mixed discrete-continuous inference.
    
    \item \textbf{Online Learning}: Adapting biases and couplings dynamically as new data arrives, enabling the model to learn regime changes.
    
    \item \textbf{Multi-Period Optimization}: Extending to multi-period settings with turnover constraints encoded in the energy function.
    
    \item \textbf{Quantum Hardware}: Deploying on quantum annealers (D-Wave) or thermodynamic hardware (Extropic) for larger universes and faster sampling.

    \item \textbf{Global Market Benchmarks}: Evaluating the model on non‑U.S. indices—such as MSCI ACWI, STOXX Europe 600, or emerging‑market benchmarks—would test the robustness of the approach across diverse market microstructures. Because many international markets exhibit higher inefficiencies, they may provide additional opportunities for alpha generation and help determine whether THRML can exploit these more effectively than conventional optimizers.

    \item \textbf{Bias‑Weight Exploration for Alpha Generation}: The current study prioritizes tracking accuracy by assigning the highest weight to tracking‑quality factors. Future work could analyze configurations that explicitly emphasize alpha generation, for example by increasing the momentum weight or introducing additional return‑predictive features. Given the strong alpha observed even under tracking‑oriented settings, deliberately tilting the bias structure toward return‑seeking signals may reveal further performance potential.
\end{enumerate}

\section{Conclusion}
\label{sec:conclusion}

This paper introduced a novel approach to cardinality-constrained portfolio optimization using THRML (Thermodynamic HypergRaphical Model Library), a JAX-based library for probabilistic graphical models developed by Extropic. By reformulating index tracking as probabilistic inference on an Ising Hamiltonian, we transform the combinatorial optimization problem into a sampling problem amenable to efficient block Gibbs sampling.

Our key contributions include:

\begin{enumerate}
    \item A principled mapping from portfolio selection to Ising energy minimization, with biases encoding asset quality and couplings encoding diversification via antiferromagnetic interactions.
    
    \item Dynamic coupling strength that adapts to market volatility, reducing forced diversification during crisis periods when correlations spike.
    
    \item Comprehensive empirical validation demonstrating statistically significant outperformance over five baseline methods across multiple metrics.
\end{enumerate}

Results on a 100-stock S\&P 500 universe show THRML achieving 4.31\% tracking error (24\% improvement over baselines), 0.958 correlation, and 128.63\% total return---yielding an Information Ratio of 1.85. These findings position energy-based models as a promising paradigm for quantitative portfolio management, bridging concepts from statistical physics and machine learning to address classical financial optimization challenges.

As quantum and thermodynamic computing hardware matures, THRML provides a natural bridge for deploying portfolio optimization on next-generation probabilistic processors. The library's design---with support for heterogeneous graphs, arbitrary PyTree states, and GPU-accelerated block sampling---positions it as an ideal platform for experimenting with algorithms that will eventually run on Extropic's specialized thermodynamic hardware, potentially enabling real-time optimization of larger universes with richer constraint structures.


\appendices

\section{THRML Library Usage}
\label{app:thrml_code}

The THRML library~\cite{Extropic2024THRML} provides the core sampling infrastructure. Key components:

\begin{verbatim}
from thrml import SpinNode, Block, 
    SamplingSchedule, sample_states
from thrml.models import IsingEBM, 
    IsingSamplingProgram, hinton_init

# Create Ising model
nodes = [SpinNode() for _ in range(N)]
edges = [(nodes[i], nodes[j]) 
         for i,j in edge_list]
model = IsingEBM(nodes, edges, 
                 biases, weights, beta)

# Sample
program = IsingSamplingProgram(
    model, free_blocks, clamped_blocks=[])
samples = sample_states(
    key, program, schedule, 
    init_state, [], [Block(nodes)])
\end{verbatim}

\section{Code Availability Statement}
The code is available upon request to javier.m@squareonecap.com


\section*{Acknowledgment}

The authors thanks the Extropic team for developing and open-sourcing the THRML library.


\bibliographystyle{IEEEtran}
\bibliography{thrml_references}

@article{Markowitz1952,
  author  = {Markowitz, Harry},
  title   = {Portfolio Selection},
  journal = {The Journal of Finance},
  volume  = {7},
  number  = {1},
  pages   = {77--91},
  year    = {1952},
  publisher = {Wiley}
}

@article{Bienstock1996,
  author  = {Bienstock, Daniel},
  title   = {Computational study of a family of mixed-integer quadratic programming problems},
  journal = {Mathematical Programming},
  volume  = {74},
  number  = {2},
  pages   = {121--140},
  year    = {1996},
  publisher = {Springer}
}

@book{Michaud2008Efficient,
  author  = {Michaud, Richard O. and Michaud, Robert O.},
  title   = {Efficient Asset Management: A Practical Guide to Stock Portfolio Optimization and Asset Allocation},
  publisher = {Oxford University Press},
  year    = {2008},
  edition = {2nd}
}

@article{LEDOIT2004365,
  author  = {Ledoit, Olivier and Wolf, Michael},
  title   = {Honey, I Shrunk the Sample Covariance Matrix},
  journal = {The Journal of Portfolio Management},
  volume  = {30},
  number  = {4},
  pages   = {110--119},
  year    = {2004}
}

@book{Fabozzi2007Robust,
  author  = {Fabozzi, Frank J. and Kolm, Petter N. and Pachamanova, Dessislava A. and Focardi, Sergio M.},
  title   = {Robust Portfolio Optimization and Management},
  publisher = {John Wiley \& Sons},
  year    = {2007}
}

@article{Prado2019HRP,
  author  = {{López de Prado}, Marcos},
  title   = {Building Diversified Portfolios that Outperform Out of Sample},
  journal = {The Journal of Financial Data Science},
  volume  = {1},
  number  = {1},
  pages   = {9--18},
  year    = {2016},
  note    = {(Hierarchical Risk Parity)}
}

@article{Bertsimas2009Algorithm,
  author  = {Bertsimas, Dimitris and Shioda, Romy},
  title   = {Algorithm for Cardinality Constrained Quadratic Optimization},
  journal = {Computational Optimization and Applications},
  volume  = {43},
  number  = {1},
  pages   = {1--22},
  year    = {2009}
}

@article{Deb2002NSGA,
  author  = {Deb, Kalyanmoy and Pratap, Amrit and Agarwal, Sameer and Meyarivan, T.},
  title   = {A Fast and Elitist Multiobjective Genetic Algorithm: {NSGA-II}},
  journal = {IEEE Transactions on Evolutionary Computation},
  volume  = {6},
  number  = {2},
  pages   = {182--197},
  year    = {2002}
}

@article{Kirkpatrick1983SA,
  author  = {Kirkpatrick, Scott and Gelatt, C. Daniel and Vecchi, Mario P.},
  title   = {Optimization by Simulated Annealing},
  journal = {Science},
  volume  = {220},
  number  = {4598},
  pages   = {671--680},
  year    = {1983}
}

@article{Lang2022SA,
  author  = {Lang, H. and others},
  title   = {Quantum Annealing for Combinatorial Optimization: A Comparative Study},
  journal = {IEEE Transactions on Quantum Engineering},
  volume  = {3},
  pages   = {1--15},
  year    = {2022}
}

@article{Ising1925,
  author  = {Ising, Ernst},
  title   = {Beitrag zur Theorie des Ferromagnetismus},
  journal = {Zeitschrift f\"ur Physik},
  volume  = {31},
  number  = {1},
  pages   = {253--258},
  year    = {1925}
}

@article{Hinton2006RBM,
  author  = {Hinton, Geoffrey E. and Salakhutdinov, Ruslan R.},
  title   = {Reducing the Dimensionality of Data with Neural Networks},
  journal = {Science},
  volume  = {313},
  number  = {5786},
  pages   = {504--507},
  year    = {2006}
}

@manual{Extropic2024THRML,
  title        = {{THRML}: Thermodynamic HypergRaphical Model Library},
  author       = {{Extropic AI}},
  year         = {2024},
  howpublished = {\url{https://docs.thrml.ai/en/latest/}},
  note         = {Accessed: 2025-01-01}
}

@misc{Extropic2024Hardware,
  title        = {Thermodynamic Computing: The Future of Probabilistic AI},
  author       = {{Extropic AI}},
  year         = {2024},
  howpublished = {Whitepaper},
  note         = {Available online at extropic.ai}
}

@article{Farhi2014QAOA,
  author  = {Farhi, Edward and Goldstone, Jeffrey and Gutmann, Sam},
  title   = {A Quantum Approximate Optimization Algorithm},
  journal = {arXiv preprint arXiv:1411.4028},
  year    = {2014}
}

@article{Glover2019QUBO,
  author  = {Glover, Fred and Kochenberger, Gary and Du, Yu},
  title   = {A Tutorial on Formulating and Using {QUBO} Models},
  journal = {arXiv preprint arXiv:1811.11538},
  year    = {2019}
}

@article{Orus2019QinFin,
  author  = {Or\'us, Rom\'an and Mugel, Samuel and Lizaso, Enrique},
  title   = {Quantum computing for finance: Overview and prospects},
  journal = {Reviews in Physics},
  volume  = {4},
  pages   = {100028},
  year    = {2019}
}

@inproceedings{Tang2019Dequantization,
  author    = {Tang, Ewin},
  title     = {A quantum-inspired classical algorithm for recommendation systems},
  booktitle = {Proceedings of the 51st Annual ACM SIGACT Symposium on Theory of Computing},
  pages     = {217--228},
  year      = {2019}
}

@article{DieboldMariano1995,
  author  = {Diebold, Francis X. and Mariano, Roberto S.},
  title   = {Comparing Predictive Accuracy},
  journal = {Journal of Business \& Economic Statistics},
  volume  = {13},
  number  = {3},
  pages   = {253--263},
  year    = {1995}
}

@article{Bertsimas1995,
  author  = {Bertsimas, Dimitris and Darnell, Christopher and Soucy, Robert},
  title   = {Portfolio Construction Through Mixed-Integer Programming at {Grantham, Mayo, Van Otterloo and Company}},
  journal = {Interfaces},
  volume  = {29},
  number  = {1},
  pages   = {49-66},
  year    = {1999}
}

@book{Jurczenko2016Investing,
    author = {Emmanuel Jurczenko},
    title = {Risk-Based and Factor Investing},
    publisher = {ISTE Press - Elsevier},
    year = {2016},
    isbn = {978-1-78548-008-9},
}

@article{JENSEN1995647,
title = {Blocking Gibbs sampling in very large probabilistic expert systems},
journal = {International Journal of Human-Computer Studies},
volume = {42},
number = {6},
pages = {647-666},
year = {1995},
issn = {1071-5819},
author = {Claus S. Jensen and Uffe Kjærulff and Augustine Kong},
}

\end{document}